\documentclass[prd,draft,preprint,showpacs,groupedaddress]{revtex4-1}
\usepackage{amsmath}
\begin{document}


\title{Emergence of space and the general dynamic equation of Friedmann-Robertson-Walker universe}


\author{Wen-Yuan Ai}
\author{Xian-Ru Hu}
\author{Hua Chen}
\author{Jian-Bo Deng}
\email[Jian-Bo Deng: ]{dengjb@lzu.edu.cn}

\affiliation{Institute of Theoretical Physics, LanZhou University, Lanzhou 730000, P. R. China}


\date{\today}

\begin{abstract}
A novel idea that the cosmic acceleration can be understood from the perspective that spacetime dynamics is an emergent phenomenon was proposed by T. Padmanabhan. The Friedmann equations of a Friedmann-Robertson-Walker universe can be derived from different gravity theories with different modifications of Padmanabhan's proposal. In this paper, a unified formula is proposed, in which those modifications can be treated as special cases when the formula is applied to different universe radii. Furthermore, the dynamic equations of a FRW universe in the $ f(R) $ theory and deformed Ho{\v{r}}ava-Lifshitz theory are investigated. The results show the validity of our proposed formula.
\end{abstract}

\pacs{04.50.-h, 04.60.-m, 04.70.Dy}

\keywords{Emergent phenomena, FRW universe, Friedmann equation}

\maketitle

\section{INTRODUCTION}
Black hole physics has provided strong hints of a deep and fundamental relationship between gravitation, thermodynamics and quantum theory. At the heart of this relationship is black hole thermodynamics \cite{bardeen1973four,hawking1975particle}, which claims that a black hole can be regarded as a thermodynamical system. But why would such a geometric object have thermodynamical characteristics? With the deep study of black holes proceeding, now it is generally believed that our spacetime is emergent. In 1995, Jacobson \cite{jacobson1995thermodynamics} first derived the Einstein equations by applying the Clausius
relation $ \delta Q=T\delta S $ on a local Rindler causal horizon. Here $ \delta S $ is the change in the entropy, while $ \delta Q $ and $ T $, respectively, represent the energy flux across the horizon and the Unruh temperature seen by an accelerating observer just inside the horizon.\par

Verlinde \cite{verlinde2011origin}, by claiming that gravity may not be a fundamental interaction but should be interpreted as an entropic force caused by changes of entropy associated with information on the holographic screen, put forward the next great step towards understanding the nature of gravity. Further, Newton's law of gravitation, the Possion equation, and in the relativistic regime the Einstein field equations, were derived with the holographic principle and the equipartition law of energy assumed. Earlier, Padmanabhan \cite{padmanabhan2010equipartition} observed that the equipartition law of energy for horizon degrees of freedom (DOF), combined with the thermodynamic relation $ S=E/2T $, leads to Newton's law of gravity.  See Refs. \cite{padmanabhan2010thermodynamical,padmanabhan2011lessons} for a review.\par

Nonetheless, in most investigations, only the gravitational field equations are treated as the equations of emergent phenomena, with the preexisting background geometric manifold assumed. If gravity can be treated as an emergent phenomenon, spacetime itself should be an emergent structure. It is obviously very hard to think that the time used to describe the evolution of dynamical variables is emergent from some pregeometric variables and the space around finite gravitational systems is emergent. Surprisingly, the conceptual difficulties disappear when one considers the emergence of spacetime in cosmology. Recently, Padmanabhan \cite{padmanabhan2012emergence} argued that our universe provides a natural setup to implement the issue that \emph{the cosmic space is emergent as cosmic time progress}. He argued that the expansion of the universe is due to the difference between the surface DOF and the bulk DOF in a region of emerged space and successfully derived the Friedmann equation of a flat Friedmann-Robertson-Walker (FRW) universe. A simple equation $ dV/dt=L_p^2\Delta N $ was proposed in Ref. \cite{padmanabhan2012emergence}, where $ V $ is the Hubble volume and $ t $ is the cosmic time.  $ \Delta N= N_{sur}-N_{bulk} $ with $ N_{sur} $ being the number of DOF on the boundary and $ N_{bulk} $ the number in the bulk.  Following Ref. \cite{padmanabhan2012emergence}, Cai generalized the derivation process to the higher $(n+1)$-dimensional spacetime. He also obtained the corresponding Friedmann equations of the flat FRW universe in Gauss-Bonnet and more general Lovelock cosmology by properly modifying the effective volume and the number of DOF on the holographic surface from the entropy formulas of static spherically symmetric black holes \cite{cai2012emergence}. The authors of \cite{yang2012emergence} took another viewpoint and derived the Friedmann equations with generalized holographic equipartition. They assumed that $ (dV/dt) $ is proportional to a function $ f(\Delta N) $. Note that the authors of Ref. \cite{cai2012emergence, yang2012emergence} only derived the Friedmann equations of the spatially flat FRW universe. In Ref. \cite{sheykhi2013friedmann}, by modifying the original proposal of Padmanabhan, Sheykhi derived the Friedmann equations of the FRW universe with any spatial curvature. He gave a new equation $dV/dt=L_p^2(\tilde{r}_A/H^{-1})\Delta N $, where $ \tilde{r}_A $ is the apparent horizon radius and $ H $ is the Hubble constant. Here $ V $ is the volume of the bulk inside the apparent horizon and $ \Delta N $ is the difference between the number of DOF on the apparent horizon and in the bulk.\par

However, different modifications relate to different gravity theories are necessary in order to obtain Friedmann equations. There are even disparate formulas for obtaining Friedmann equations of a flat FRW universe and for obtaining the equations of one with any spatial curvature. We believe that if the idea suggested by Padmanabhan is right, there should be a unified formula describing the emergence. In this paper, we propose a general formula and argue that the modifications made by Cai and Sheykhi can be treated as special cases with the formula being applied to different universe radii when the general relationship between the number of DOF on the surface and the entropy of the horizon \cite{tu2013emergence} is adopted. With our proposed formula, further study of the dynamic equations of a FRW universe in the $ f(R) $ theory and the deformed Ho{\v{r}}ava-Lifshitz (HL) theory can be carried out. The paper is organized as follows: In \mbox{Sec. \uppercase\expandafter{\romannumeral3}}, we present our formula and show in what cases the formula would be reduced to those of Padmanabhan, Cai and Sheykhi. In \mbox{Sec. \uppercase\expandafter{\romannumeral3}}, the $ f(R) $ theory and deformed HL theory are investigated. \mbox{Section \uppercase\expandafter{\romannumeral4}} is for conclusions and discussions. We take $ k_B=1 $ through all this paper.

\section{GENERAL FORMULA FOR THE EMERGENCE OF COSMIC SPACE}
First, let us recall Padmanabhan and the others' work \cite{padmanabhan2012emergence, cai2012emergence, sheykhi2013friedmann, tu2013emergence}. Padmanabhan notices that for a pure de Sitter universe with Hubble constant H, the holographic principle can be expressed in terms of the form \cite{padmanabhan2012emergence}
\begin{equation}
\label{1} N_{sur}=N_{bulk},
\end{equation}
where $N_{sur}$ denotes the number of DOF on the spherical surface of Hubble radius $ H^{-1} $, $ N_{sur}=4\pi H^{-2}/L_p^2 $, with $ L_p $ being the Planck length, while the bulk DOF $ N_{bulk}=\left|E\right|/(1/2)T $. Here $ \left|E\right|=\left|\rho+3p\right|V $, is the Komar energy and the horizon temperature $ T=H/2\pi $. For the pure de Sitter universe, by substituting $ \rho=-p $ into Eq. \eqref{1}, the standard result $ H^2=8\pi L_p^2\rho/3 $ is obtained.\par
One can get $ \left|E\right|=(1/2)N_{sur}T $ from the equality \eqref{1}, which is the standard equipartition law. Since it relates the effective DOF residing in the bulk to the DOF on the boundary surface, Padmanabhan called it \emph{holographic equipartition}. However, as shown by lots of astronomical observations, our real universe is not a pure de Sitter but asymptotically de Sitter universe. Padmanabhan further considers that, for our real universe, the expansion is caused by the emergence of space and relates to the difference $ \Delta N=N_{sur}-N_{bulk} $. He proposes a simple equation as \cite{padmanabhan2012emergence}
\begin{equation}
\label{2} \frac{dV}{dt}=L_P^2\Delta N.
\end{equation}
This suggests that the expansion of the universe is being driven towards holographic equipartition. It can be treated as the thermodynamical process on the horizon and therefore comes to be an emergent phenomenon. Putting the above definition of each term, one obtains
\begin{equation}
\label{3} \frac{\ddot{a}}{a}=\frac{-4\pi L_p^2}{3}(\rho+3p).
\end{equation}
This is the standard dynamical equation for a FRW universe in general relativity. With the help of the continuity equation $ \overset{.}{\rho}+3H(\rho +p)=0 $, one gets the standard Friedmann equation
\begin{equation}
\label{4} H^2+\frac{k}{a^2}=\frac{8\pi L_p^2}{3}\rho,
\end{equation}
where $ k $ is an integration constant which can be interpreted as the spatial curvature of the FRW universe. Here the cosmological constant term has already been taken into account in the Komar energy. In fact, in order to have the asymptotic holographic equipartition, Padmanabhan takes $(\rho+3p)<0$. This implies the existence of dark energy is necessary. Put another way, \emph{the existence of a cosmological constant in the universe is required for asymptotic holographic equipartition} \cite{padmanabhan2012emergence}.\par
Then Cai generalized the above derivation process to the higher (n+1)-dimensional case with $ n>3 $. In this case, the number of DOF on the holographic surface is given by \cite{verlinde2011origin}
\begin{equation}
\label{5} N_{sur}=\alpha A/L_p^{n-1}.
\end{equation}
Here $ A=n\Omega_n/H^{n-1} $ and $ \alpha =(n-1)/2(n-2) $ with $ \Omega_n $ being the volume of an n-sphere with unit radius. In this case, Cai made a minor modification for the proposal \eqref{2} as \cite{cai2012emergence}
\begin{equation}
\label{6} \alpha\frac{dV}{dt}=L_p^{n-1}(N_{sur}-N_{bulk}),
\end{equation}
where the volume $ V=\Omega_n/H^n $. Here the bulk DOF remains the same i.e. $ N_{bulk}=-2E/T $ when we only consider the accelerating phase with $ (n-2)\rho+np<0 $ and the bulk Komar energy is \cite{cai2010friedmann}
\begin{equation}
\label{7} E_{Komar}=\frac{(n-1)\rho+np}{n-2}V.
\end{equation}
With the help of the continuity condition, after some simple algebra, the Friemann equation can be obtained finally.\par
Moreover, for Gauss-Bonnet gravity and Lovelock gravity, the volume increase was modified as follows \cite{cai2012emergence}
\begin{equation}
\label{8} \frac{d\tilde{V}}{dt}=\frac{1}{(n-1)H}\frac{d\tilde{A}}{dt},
\end{equation}
where $ \tilde{A}/L_p^{n-1}=4S $, $ S $ is the entropy formula of black holes modified in Gauss-Bonnet gravity and Lovelock gravity. Further, the number of DOF on the surface must be taken special forms until the Friedmann equations of the spatially flat FRW universe can be obtained \cite{cai2012emergence}.\par
To get the Friedmann equations of the FRW universe with any spatial curvature in Gauss-Bonnet and Lovelock gravities, Sheykhi proposed \cite{sheykhi2013friedmann}
\begin{equation}
\label{9} \alpha \frac{d\tilde{V}}{dt}=L_p^{n-1}\frac{\tilde{r}_A}{H^{-1}}(N_{sur}-N_{bulk}),
\end{equation}
where $ \tilde{r}_A=1/\sqrt{H^2+k/a^2} $ is the apparent horizon radius, $ H $ is the Hubble constant and $ V $ is respect to the volume of the bulk inside the apparent horizon, $ N_{sur} $ and $ N_{bulk} $ are respectively the number of DOF on the apparent horizon and in the bulk.\par
Even though we can derive Friedmann equations from special forms of $ N_{sur} $, it is suspect that $ N_{sur} $ must be chosen in such complicated forms. Compared with this, the general relationship between the number of DOF on the surface and the entropy of the horizon proposed by the authors of Ref. \cite{tu2013emergence} is more natural, i.e.
\begin{equation}
\label{10} N_{sur}=4S.
\end{equation}
With this relationship adopted, we propose a more general equation
\begin{equation}
\label{11} \frac{\alpha}{H(n-1)}\frac{dN_{sur}}{dt}=N_{sur}-N_{bulk}.
\end{equation}
When $ n=3 $, we have $ \alpha=1 $, $ S=A/4L_p^2 $, $ dN_{sur}/dt=dA/(L_p^2dt) $ and $dV/dt=dA/2Hdt$, thus Eq.\eqref{1} proposed by Padmanabhan is recovered immediately.
By substituting $ N_{sur}=4S=\tilde{A}/L_p^{n-1} $, $ d\tilde{V}/dt=d\tilde{A}/(n-1)Hdt $ into Eq. \eqref{11} and applying Eq. \eqref{11} to the Hubble radius, we can obtain Eq. \eqref{6}, which is the modification given by Cai. Similarly, if we apply Eq. \eqref{11} to the apparent horizon radius, in this case, $ d\tilde{V}/dt=\tilde{r}_Ad\tilde{A}/(n-1)dt $, substituting it into Eq. \eqref{11}, Eq. \eqref{9} proposed by Sheykhi is obtained immediately.\par
We have shown the universality of Eq. \eqref{11}. When we consider the (3+1)-dimensional Einstein gravity, it gets back to Padmanabhan's original equation. However, Eq. \eqref{11} can also describe the evolution of a universe in higher (n+1)-dimensional Einstein gravity, Gauss-Bonnet gravity or more general Lovelock gravity. Our proposed equation is the general form that obtains Eq. \eqref{6}  when applied to the Hubble radius and Eq. \eqref{9} when applied to the apparent radius.
\section{THE DYNAMIC EQUATIONS OF A FRW UNIVERSE IN THE $ f(R) $ THEORY AND DEFORMED HL THEORY}
Up to now, we have argued that Eq. \eqref{11} is the general formula which can be used to investigate the emergence of spacetime in cosmology. Now we will use it to push further study on the dynamic equations of a FRW universe in the $ f(R) $ theory and the deformed HL theory. For simplicity, we only consider the case when Eq. \eqref{11} is applied to the Hubble sphere. \par
Let us first consider the emergence of space in $ f(R) $ theory. $ f(R) $ gravity was first proposed in 1970 by H. A. Buchdahl \cite{buchdahl1970non}, as a type of modified gravity theory that generalizes Einstein's general relativity. It is actually a family of theories, each one defined by a different function of the Ricci scalar. When $f(R)=R$, it goes back to general relativity. In $f(R)$ theory, We take the Komar energy generally as \cite{padmanabhan2004entropy}
\begin{equation}
\label{12} \left|E\right|=2\left|\int_{\Sigma}\left(T_{\mu\nu}-\frac{1}{2}Tg_{\mu\nu}\right)U^{\mu}U^{\nu}\right|=\left|\rho+3p\right|V,
\end{equation}
where $ T_{\mu\nu} $ is the total energy-momentum tensors which contains two parts, the energy-momentum tensor of matter and gravity. $ U^{\mu}$, $U^{\nu} $ are the components of $ U^{a} $, which is the tangent vector of isotropic observers' wordline in a FRW universe. One can obtain the total energy density $ \rho $ and total pressure $ p $ through $ T_{\mu\nu} $. The black hole entropy is given by \ \cite{wald1993black}
\begin{equation}
\label{13} S=\frac{f^\prime A}{4L_p^{n-1}}.
\end{equation}
From relation \eqref{10}, we get
\begin{equation}
\label{14} N_{sur}=\frac{f^\prime A}{L_p^{n-1}},
\end{equation}
where $ f^\prime $ denotes $ f(R) $ derivatives taken with respect to $ R $. Substituting $ V=\Omega_n/H^n $ into Eq. \eqref{7} and using $ N_{bulk}=-2E/T $, we have
\begin{equation}
\label{15} N_{bulk}=\frac{-4\pi \Omega_n[(n-2)\rho +np]}{(n-2)H^{n+1}}.
\end{equation}
According to Eq. \eqref{11}, we obtain the dynamic equation
\begin{equation}
\label{16} H^2+\frac{n-1}{2(n-2)}\dot H=\frac{-4\pi L_p^{n-1}[(n-2)\rho+np]}{n(n-2)f^\prime}.
\end{equation}
This is the formal dynamic equation of the FRW universe we got from the emergence of space in the (n+1)-dimensional $ f(R) $ theory. When $ n=3 $, we have
\begin{equation}
\label{17} H^2+\dot{H}=\frac{-4\pi L_p^2(\rho+3p)}{3f^\prime}.
\end{equation}\par
We know that the total energy-momentum tensor $ T_{\mu\nu} $ cannot be determined if we do not give the specific form of the theory. Now, we will determine that in the FRW model. By assuming the background spacetime is spatially homogeneous and isotropic, one can find the FRW metric
\begin{equation}
\label{18} ds^2=-dt^2+a^2(t)\left(\frac{dr^2}{1-kr^2}+r^2d\Omega_{n-2}^2\right),
\end{equation}
where $d\Omega_{n-2}^2$ denotes the line element of an (n-1)-dimensional unit sphere and the spatial curvature constants $ k=0,1,-1 $ correspond to a flat, closed, and open universe, respectively. Based on the cosmological principle, we take the components of $ T_{\mu\nu} $ as
\begin{equation}
\label{19} T_{00}=\rho(t),T_{0i}=0,T_{ij}=a^2(t)\delta_{ij}p(t),
\end{equation}
where $ i,j $ run over $ 1,2,...,n-1$.\par
As in Ref. \cite{tu2013emergence}, we can use the Einstein-Hilbert (EH) action to determine the total energy-momentum tensor. In the $ f(R) $ theory the EH action takes the form
\begin{equation}
\label{20} S=\int d^4x\sqrt{-g}\left(f(R)+2\kappa^2L_{m}\right),
\end{equation}
where $ \kappa^2=8\pi G $. $ L_{m} $ is the Lagrangian density of matter in this theory. After using the variational principle $ \delta S=0 $, one can obtain
\begin{equation}
\label{21} R_{\mu\nu} f^\prime -\frac{1}{2}g_{\mu\nu}f(R)+g_{\mu\nu}\nabla^2f^\prime-\nabla_\mu\nabla_\nu f^\prime=\kappa^2 T^{(m)}_{\mu\nu},
\end{equation}
where $ T^{(m)}_{\mu\nu} $ is the energy-momentum tensor of the matter. Adopting the Einstein tensor $ G_{\mu\nu}=R_{\mu\nu}-Rg_{\mu\nu}/2 $, one has
\begin{equation}
\label{22} G_{\mu\nu}f^\prime=\kappa^2T^{(m)}_{\mu\nu}+\frac{f(R)-Rf^\prime}{2}g_{\mu\nu}+\nabla_\mu\nabla_\nu f^\prime-g_{\mu\nu}\nabla^2f^\prime.
\end{equation}
With the energy-momentum tensor of the gravity caused by higher-order derivatives defined as
\begin{equation}
\label{23} T_{\mu\nu}^{(g)}=\frac{1}{f^\prime}\left(\frac{f(R)-Rf^\prime}{2}g_{\mu\nu}+\nabla_\mu\nabla_\nu f^\prime-g_{\mu\nu}\nabla^2f^\prime\right),
\end{equation}
one can arrive at
\begin{equation}
\label{24} G_{\mu\nu}=\kappa^2\left(\frac{1}{f^\prime}T^{(m)}_{\mu\nu}+\frac{1}{\kappa^2}T^{(g)}_{\mu\nu}\right)\equiv\kappa^2T_{\mu\nu}.
\end{equation}
With the matter to be perfect fluid assumed, according to the form Eq. \eqref{19} and
\begin{equation}
\label{25} T^{(m)}_{00}=\rho_m(t),\ T^{(g)}_{00}=\rho_g(t),\ T^{(m)}_{ij}=a^2(t)\delta_{ij}p_m(t),\ T^{(g)}_{ij}=a^2(t)\delta_{ij}p_g(t),
\end{equation}
one can get
\begin{equation}
\label{26} \rho=\frac{1}{f^\prime}\left[\rho_m+\frac{1}{\kappa^2}\left(\frac{Rf^\prime-f}{2}-3Hf^{\prime\prime}\dot{R}\right)\right],
\end{equation}
\begin{equation}
\label{27} p=\frac{1}{f^\prime}\left[p_m+\frac{1}{\kappa^2}\left(-\frac{Rf^\prime-f}{2}+f^{\prime\prime\prime}\dot{R}^2+f^{\prime\prime}\ddot{R} +2Hf^{\prime\prime}\dot{R}\right)\right]
\end{equation}
in a FRW universe, where $ \dot{R} $ denotes $ f(R) $ derivatives taken with respect to $ t $. By substituting Eqs. \eqref{26} and \eqref{27} into Eq. \eqref{16}, we finally obtain the dynamic equation of a FRW universe from the idea of the emergence of space. When $ f(R)=R $, it has $ f^\prime=1 $, $ \rho_g=p_g=0 $. Thus, from Eq. \eqref{17} the standard dynamic equation of the FRW universe in general relativity
\begin{equation}
\label{28} H^2+\dot{H}=\frac{-4\pi L_p^2(\rho_m+3p_m)}{3}
\end{equation}
is obtained, showing consistency.\par
We regard Eq. \eqref{16} or Eq. \eqref{17} as the modified dynamic equation of a FRW universe with a global correction from the idea of the emergence of space.\par
Next, we will consider the emergence of space in deformed HL gravity. Motivated by Lifshitz theory in solid state physics, Ho{\v{r}}ava proposed a new gravity theory at a Lifshitz point \cite{hovrava2009quantum, hovrava2009membranes, hovrava2009spectral}. The theory has manifest 3-dimensional spatial general covariance and time reparametrization invariance. This is a non-relativistic renormalizable theory of gravity and
recovers the four dimensional general covariance only in an infrared limit. Thus, it may be regarded as a UV complete candidate for general relativity. In the deformed HL gravity, we also take the form of the Komar energy as in the f(R) theory. The entropy has the form \cite{myung2010entropy}
\begin{equation}
\label{29} S=\frac{A}{4L_p^{n-1}}+\frac{\pi}{\omega}\ln\frac{A}{4 L_p^{n-1}},
\end{equation}
where the parameter $\omega=16\mu^2/\kappa^2$. The entropy/area relation has a logarithmic term, which is a characteristic of HL gravity theory. However, as the parameter $\omega \rightarrow \infty$, it goes back to the one in Einstein gravity. Using the relation in Eq. \eqref{10}, one gets
\begin{equation}
\label{30} N_{sur}=\frac{A}{L_p^{n-1}}+\frac{4\pi}{\omega}\ln\frac{A}{4 L_p^{n-1}}.
\end{equation}
$N_{bulk}$ is as same as Eq. \eqref{15}. By substituting every term into Eq. \eqref{11}, we get
\begin{equation}
\label{31}
\begin{aligned}
H^2+\frac{n-1}{2(n-2)}\dot{H}=&-\frac{4\pi L_p^{n-1}[(n-2)\rho+np]}{n(n-2)}\\
&-\frac{2(n-1)\pi L_p^{n-1}H^{n-1}\dot{H}}{n(n-2)\omega\Omega_n}\\
&-\frac{4\pi L_p^{n-1}H^{n+1}}{n\omega\Omega_n}\ln\frac{n\Omega_n}{4L_p^{n-1}H^{n-1}}.\\
\end{aligned}
\end{equation}
When $n=3$, then $ \Omega=4\pi/3 $. Thus, we have
\begin{equation}
\label{32}H^2+\dot{H}=-\frac{4\pi L_p^2(\rho+3p)}{3}-\frac{L_p^2H^2\dot{H}}{\omega}-\frac{L_p^2H^4}{\omega}\ln\frac{\pi}{L_p^2H^2}.
\end{equation}
Now, we have proved that the standard dynamic equation of the FRW universe in general relativity can be recovered when the parameter $ \omega\rightarrow\infty $.\par
We define the total energy and pressure of the universe in HL theory as \cite{wang2009thermodynamics, cao2010clausius}
\begin{equation}
\label{33} \rho=\rho_m+\rho_A+\rho_k+\rho_{dr},\ p=p_m+p_A+p_k+p_{dr},
\end{equation}
where the matter is assumed to be a perfect fluid. The cosmological constant term, the curvature term, and the dark radiation term are \cite{cao2010clausius}
\begin{align}
&\rho_A=-p_A=-\frac{3\kappa^2\mu^2{\Lambda_W}^2}{8(3\lambda-1)}, \\
&\rho_k=-3p_k=\frac{3k}{4(3\lambda-1)a^2}\left(\kappa^2\mu^2{\Lambda_W}-8\mu^4(3\lambda-1)+\frac{8}{\kappa^2}(3\lambda-1)^2\right),\\
&\rho_{dr}=3p_{dr}=\frac{3\kappa^2\mu^2}{8(3\lambda-1)}\frac{k^2}{a^4}.
\end{align}\par
At last, with the speed of light and Newton's constant in the IR limit given by \cite{myung2010entropy}
\begin{equation}
\label{37} c^2=\frac{\kappa^2\mu^4}{2},\ G=\frac{\kappa^2}{32\pi c},\ \lambda=1,
\end{equation}
one can get the specific form of entropy Eq. \eqref{29}. From this, with Eqs. \eqref{31} and \eqref{33}, one can finally get the complete dynamic equation of a FRW universe. Here, as in the case of $f(R)$ theory, Eq. \eqref{31} or Eq. \eqref{32} is the modified dynamic equation in HL theory from the global viewpoint of emergence. \par
So far, we have shown that Eq. \eqref{11} can be used to derive dynamic equations of a FRW universe in the $f(R)$ gravity and deformed HL gravity. Now, we would like to say that our equation can be treated as the unified dynamic equation of a FRW universe in all the gravity theories involved in this paper.\par

Note that the dynamic equations for a FRW universe in $ f(R) $ and the deformed HL theory obtained here are different from those obtained in Ref. \cite{tu2013emergence}. Since we have argued that Eq. \eqref{11} is a more general formula of describing the cosmic emergence, we believe that it is better to use Eq. \eqref{11} than Eq. \eqref{2} to derive the dynamic equation of a FRW universe while applying the idea of Padmanabhan to the gravity theories beyond general relativity.
\section{DISCUSSION AND CONCLUSION}
In this paper, we investigated the novel idea proposed by Padmanabhan \cite{padmanabhan2012emergence}, which states that the emergence of space and the universe's expansion can be understood by calculating the difference between the number of DOF on the surface and in the bulk. We also summarized the work done by Cai, Sheykhi and others \cite{cai2012emergence, sheykhi2013friedmann, tu2013emergence}. Since different modifications of Padmanabhan's proposal relate to different gravity theories until finally the Friedmann equations are obtained, which seems unreasonable, we proposed a general formula Eq.\eqref{11} which can be reduced to the different modified ones in different cases. When we consider the (3+1)-dimensional Einstein gravity, it gets back to Padmanabhan's original Eq. \eqref{2}. We argued that the equations proposed by Cai and Sheykhi are, respectively, special cases of our proposed equation being applied to the Hubble radius and apparent radius.\par
 We also pushed for further research on the dynamic equations of a FRW universe in $ f(R) $ theory and deformed HL theory using the proposed formula. With the idea proposed by Padmanabhan, we finally obtained the modified dynamic equations from the perspective of emergent phenomena. The resulting equations show a strong consistency with the standard dynamic equations of the FRW universe in general relativity when $n=3$, $f(R)=R $ and $ \omega\rightarrow\infty $. These results imply that our formula is not only valid for general relativity, higher dimensional Einstein gravity, Gauss-Bonnet gravity, and Lovelock gravity, but is also valid for $ f(R) $ gravity and deformed HL gravity. Therefore, Eq. \eqref{11} can be treated as the unified formula governing the emergence of cosmic space and can be widely used to study the dynamic equations of FRW universes. Even though we have shown the universality of Eq. \eqref{11}, we have not found the origin of our proposed formula. However, since Eq. \eqref{11} is valid for all the gravity theories mentioned here, we still believe it has a deep one. This, surely, is worthy of further investigation.

\bibliography{refrence}

\end{document}